\UseRawInputEncoding
\documentclass[aps,prd,twocolumn,UFT8]{revtex4}
\pdfoutput=1

\usepackage{mathrsfs}
\usepackage{amsmath,array}
\usepackage{bm}
\usepackage{float}
\usepackage{hyperref}
\usepackage{cleveref}
\usepackage{bbding}
\usepackage{pifont}
\usepackage{txfonts}
\usepackage{color}
\usepackage{ulem}
\usepackage{appendix}

\newcommand{\bea}{\begin{eqnarray}}
\newcommand{\eea}{\end{eqnarray}}
\newcommand{\beq}{\begin{equation}}
\newcommand{\eeq}{\end{equation}}
\newcommand{\nn}{\nonumber}
\def\/{\over}

\begin{document}

\title{\bf Interatomic interaction of two ground-state atoms in vacuum:  contributions of vacuum fluctuations and  radiation reaction}
\author{Wenting Zhou$^{1,2}$, Shijing Cheng$^{3}$, and Hongwei Yu$^{3}$\footnote{Corresponding author: hwyu@hunnu.edu.cn}}
\affiliation{$^{1}$ Department of Physics, School of Physical science and Technology, Ningbo University, Ningbo, Zhejiang 315211, China\\
$^{2}$ Center for Nonlinear Science, Ningbo University, Ningbo, Zhejiang 315211, China\\
$^{3}$ China Key Laboratory of Low Dimensional Quantum Structures and Quantum Control of Ministry of Education and Synergetic Innovation Center for Quantum Effect and
Applications, \\Hunan Normal University, Changsha, Hunan 410081, People¡¯s Republic of China
}

\begin{abstract}

We generalize the formalism proposed by Dalibard, Dupont-Roc and Cohen-Tannoudji [the DDC formalism] to the fourth order of the coupling constant, which can be used to study the interatomic interaction of two ground-state atoms coupled with the vacuum scalar fields. We show that the interatomic potential can be attributed to the joint effect of both vacuum fluctuations and the radiation reaction of atoms.  Remarkably, the formulae we derived for  the contributions of vacuum fluctuations and the radiation reaction to the interatomic potential upon which future research on fourth-order effects in particular circumstances can be based differ from those in the existing literature [Phys. Rev. D 95, 085014 (2017)].
\end{abstract}

\maketitle

\section{Introduction}

The radiative properties of atoms, the spontaneous emission for instance, have been a long-standing and fascinating topic in  the  research of the atom-field interaction.  So far   vacuum field fluctuations~\cite{Weisskopf} and atom's radiation reaction~\cite{Ackerhalt73}, or a combination of them~\cite{Ackerhalt73,Senitzky73,Milonni75,Milonni88}, have been put forward as possible physical mechanisms  underlying  the spontaneous emission. The ambiguity in the physical explanation arises from an  indetermination in the separation of effects of the vacuum field fluctuations and  the atom's radiation reaction as a result of the freedom of choice in the ordering of the operators of the field and the atoms~\cite{Milonni73}, such that distinct contributions of the vacuum fluctuations and the radiation reaction to the spontaneous emission of atoms may not possess an independent physical meaning. The controversy remained until  Dalibard, Dupont-Roc and Cohen-Tannoudji[DDC]  proposed that one can get rid of the ambiguity by adopting a preferred operator ordering, i.e., the symmetric operator ordering between the variables of the atoms and the field,  so that the Hamiltonians of the contributions of the vacuum fluctuations and the atom's radiation reaction are respectively Hermitian and thus possess independent physical meanings~\cite{DDC82,DDC84}. This approach  was then widely exploited to study the roles of the field fluctuations and the atom's radiation reaction in the atom-field interaction.

The studies in  the average variation rate of energy of an atom in interaction with various kinds of fluctuating vacuum quantum fields by the use of the DDC formalism, such as the scalar field, the electromagnetic field, the Dirac field and the Rarita-Schwinger field, provide clear physical explanations for the stability in vacuum of an inertial ground-state atom, and the spontaneous excitation of a non-inertial atom in a flat background~\cite{Meschede90,Audretsch94,Audretsch952,Zhu06,Yu06,Zhou12,Li14} and a static atom in a gravitational field~\cite{Yu071}, as well as an understanding of the Unruh effect from a different perspective~\cite{Zhu07,Zhou08}. 

The DDC formalism was also utilized to investigate the energy shifts of an atom in interaction with fluctuating vacuum quantum fields in various circumstances.  It is discovered that the energy shifts can be greatly modified by the  non-inertial motion of an atom in vacuum~\cite{Audretsch95,Audretsch952,Passante98,Rizzuto07,Rizzuto09,Rizzuto092,Zhu10,Rizzuto11}. In particular, when the atom is accelerated in an unbounded space, the contributions of the vacuum fluctuations to the energy shifts  are obviously modified while the contributions of the atom's radiation reaction remain unaltered as compared to that of an inertial atom~\cite{Audretsch95,Audretsch952,Passante98}. However,  if the acceleration occurs  in a bounded space, such as near a boundary or in a cavity, both the contributions of the vacuum fluctuations and the  atom's radiation reaction are modified by the noninertial motion of the atom~\cite{Rizzuto07,Rizzuto09,Rizzuto092,Zhu10,Rizzuto11}. Besides, the energy shifts of an atom in a thermal bath in equilibrium~\cite{Tomazelli03,Zhu092} or out of thermal equilibrium are also examined~\cite{Zhou14}. All these studies are concerned with  a single atom.

When two atoms are considered, an interatomic interaction potential arises as a result of the fact that one atom interacts with the fluctuating  field and a radiative field is induced, which then acts on the other atom, and vice versa. The simplest interatomic interaction, from the viewpoint of quantum electrodynamics, is the resonance coupling of two identical atoms  with one  in an excited state~\cite{Craig98}. For two identical atoms in the symmetric/antisymmetric entangled state, 
it has been shown with the perturbation theory that the transitions between these states [the symmetric/antisymmetric entangled states] and the states with both atoms in their ground states or excited states are a second order perturbation effect, and so is the resonance interatomic potential. 
Recently, the resonance interaction of two identical atoms has also been studied by the use of the DDC formalism. It is discovered that in sharp contrast to the radiative properties of a single atom in interaction with the scalar or electromagnetic fields, such as the energy shifts or the average variation rates of  energy of the atom, the resonance interatomic energy of two identical atoms in the symmetric/antisymmetric entangled state is wholly caused by the radiation reaction of the atoms but irrelevant to the vacuum fluctuations of quantum fields~\cite{Rizzuto16,Zhou163,Zhou18,Zhou183}; while the energy shifts and the average variation rate of energy of a single atom in interaction with the vacuum scalar or the electromagnetic field are a result of the joint effect of both  the vacuum fluctuations and the atom's radiation reaction. Recently, the roles of the vacuum fluctuations and the radiation reaction in the transitions between the symmetric/antisymmetric entangled state and other eigenstates [with both atoms in their ground or excited states] of the two-atom system are also discussed~\cite{Menezes15,Menezes16,Menezes162,Liu18,Cai18,Zhou201,Zhou202}.

The investigations mentioned above are about two atoms in an unfactorizable state. The interatomic interaction between two atoms in a factorizable state, two atoms in their ground states and two atoms with at least one in its excited state,  for instance,  is however more of a commonplace. In these cases, the leading  interatomic interaction are no longer second order effects but rather fourth order ones, and are thus more difficult to deal with.
For two ground-state atoms in interaction with vacuum quantum electromagnetic fields, Casimir and Polder investigated the interatomic potential by using the stationary-state perturbation theory~\cite{Casimir48}. And the interatomic interaction of two atoms with at least one  in an excited state, which is a little more complicated 
since an exchange of real photons may be  involved, has also been extensively studied and many interesting results are obtained~\cite{Power94,Power95,Berman15,Milonni15,Donaire15,Donaire16,Donaire162,Donaire17,Jentschura17}.
Let us note that the interatomic interaction between two ground-state atoms can also be analyzed using the DDC formalism where the contributions of vacuum fluctuations and the radiation reaction are distinctively separated. In this regard,
recently, by generalizing the DDC formalism to the fourth order, Marino \textit{et al} considered the interatomic potential and the Casimir-Polder force of a pair of ground-state atoms in two cases, i.e. two atoms in synchronized uniform acceleration in vacuum and two static atoms immersed in a thermal bath, and concluded that both thermal and nonthermal features associated with two atoms in uniform acceleration can be probed through the Casimir-Polder force between two accelerating atoms~\cite{Marino14}. As the first attempt in interpreting the interatomic potential as a result of vacuum fluctuations and  the atomic radiation reaction, a toy model of atoms in the monopole interaction with the vacuum quantum scalar field is used. The fourth order DDC formalism was later generalized from the flat Minkowski spacetime to the curved Schwarzschild spacetime, and the thermal and nonthermal behaviors of the Casimir-Polder interaction of two ground-state atoms coupled with vacuum scalar fields outside a Schwarzschild black hole were analyzed~\cite{Menezes17}. To the best of our knowledge, these  are the only works dealing the interatomic potential with the fourth order DDC formalism so far. However, in both these two papers, a detailed derivation of the DDC formalism is absent, and moreover, the formula for the contributions of the atomic radiation reaction to the interatomic potential  was not given in Ref.~\cite{Marino14}, while  that given in Ref.~\cite{Menezes17} does not seem to be correct. Actually, as we will demonstrate later, the generalization of the DDC formalism  to the fourth order is not as  straightforward as one may think  but rather tricky.

Therefore, the purpose of the present paper is to give a detailed derivation for the contributions of the vacuum fluctuations and the atomic radiation reaction to the interatomic potential of two ground-state atoms in interaction with quantum scalar fields in vacuum, obtaining general  and detailed  formulae upon which  the future research  on fourth-order effects in particular circumstances can be based. Here it is worth stressing that although here the interatomic potential is considered  in a simple toy model of atoms with the induced monopole interacting with the vacuum quantum scalar field, a  generalization to a more realistic model in which  atoms interact with the quantum electromagnetic field via the induced dipole interaction is straightforward as it will become clear in the following derivation. We work in the Heisenberg picture, which allows an easy comparison of the quantum-mechanical and classical concepts. Our paper is organized as follows. In section II, 
we first obtain and then solve the Heisenberg equations of motion of the dynamical variables of atoms and the field. Then we split each solution into a free part and a source part. In section III, we consider order by order the average variation rate of the Hamiltonian of the two atoms in terms of the contributions of the free field and the source field. To ensure the Hermitian property of the operators corresponding to the contributions of the vacuum fluctuations and the atomic radiation reaction, a symmetric operator ordering~\cite{DDC82,DDC84} between the variables of the atoms and the field is adopted. In section IV, by treating the two atoms as a whole, we derive the effective Hamiltonians of the contributions of vacuum fluctuations and the atomic radiation reaction. In section V, we evaluate the expectation values of the effective Hamiltonians over the state of the two-atom system, which then give rise to
to the contributions of the vacuum fluctuations and the atomic radiation reaction to the interatomic potential. It turns out that 
the contributions of the atomic radiation reaction derived in this paper is quite different from that obtained in Ref.~\cite{Menezes17}. Note also that the formula for the contributions of the atomic radiation reaction has not been reported in Ref.~\cite{Marino14}. 
We give a brief summary for our work in section VI.

\section{Time evolution of the dynamical variables of the atoms and the field}

Suppose that a pair of two-level atoms are in interaction with the fluctuating scalar field in vacuum. We label the two atoms respectively by $A$ and $B$, and their transition frequencies  by $\omega_A$ and $\omega_B$. The two atoms are assumed to move synchronously along stationary trajectories.  Thus their proper time is the same which we will denote by $\tau$. Accordingly, the trajectories of the atoms are denoted by $x_A(\tau)$ and $x_B(\tau)$. As the two atoms are coupled to the fluctuating fields in vacuum, there is an induced  interaction between them which is described by the interatomic potential. We aim to derive the contributions of vacuum fluctuations and the atomic radiation reaction to the interatomic potential. Our derivation will be carried out in the Heisenberg picture.

The Hamiltonian of the two atoms, $H_S(\tau)$, in Dicke's notation~\cite{Dicke54}, can be written as
\beq
H_S(\tau)=\omega_AR_{3}^{A}(\tau)+\omega_BR_{3}^{B}(\tau)\;,\label{H_S}
\eeq
where $R_{3}^{\xi}=\frac{1}{2}(|e_{\xi}\rangle\langle e_{\xi}|-|g_{\xi}\rangle\langle g_{\xi}|)$, $(\xi=A,B)$ with $|g_{\xi}\rangle$ and $|e_{\xi}\rangle$ being the ground state and the excited state of atom $\xi$ with energy $-\frac{1}{2}\omega_{\xi}$ and $+\frac{1}{2}\omega_{\xi}$ respectively. The Hamiltonian of the scalar field with respect to the proper time, $\tau$, is given by
\beq
H_F(\tau)=\int d^3{\mathbf{k}}\;\omega_{\mathbf{k}}a^{\dag}_{\mathbf{k}}(t)a_{\mathbf{k}}(t)\frac{dt}{d\tau}\;,
\eeq
where $\mathbf{k}$ denotes the wave vector of the  the scalar field modes, and $a^{\dag}_{\mathbf{k}}(t)$ and $a_{\mathbf{k}}(t)$ are respectively the creation and annihilation operators with momentum $\mathbf{k}$. The Hamiltonian that describes the  interaction between the atoms  and the scalar field, $H_I(\tau)$, which is assumed to be weak,  can be expressed as
\beq
H_I(\tau)=\mu R_{2}^{A}(\tau)\phi(x_A(\tau))+\mu R_{2}^{B}(\tau)\phi(x_B(\tau))\;,
\eeq
where $\mu$ is a small coupling constant,
\beq
R_{2}^{\xi}(\tau)=\frac{i}{2}(R_{-}^{\xi}(\tau)-R_{+}^{\xi}(\tau))\label{sigma-2}
\eeq
is the monopole operator of atom $\xi$ with $R_+^{\xi}=|e_{\xi}\rangle\langle g_{\xi}|$ and $R_{-}^{\xi}=|g_{\xi}\rangle\langle e_{\xi}|$ being the atomic raising and lowering operators, and $\phi(x(\tau))$ is the scalar field operator,
\beq
\phi(t,\mathbf{x})=\int d^3{\mathbf{k}}\;g_{\mathbf{k}}[a_{\mathbf{k}}(t)e^{i\mathbf{k}\cdot\mathbf{x}}+a^{\dag}_{\mathbf{k}}(t)e^{-i\mathbf{k} \cdot \mathbf{x}}]
\eeq
with $g_{\mathbf{k}}=[(2\pi)^3 2\omega_{\mathbf{k}}]^{-1/2}$. Then the total Hamiltonian of the ``atoms+field" system is given by the sum of the above three Hamiltonians:
\bea
H(\tau)&=&\omega_AR_{3}^{A}(\tau)+\omega_BR_{3}^{B}(\tau)+\int d^3{\mathbf{k}}\omega_{\mathbf{k}}a^{\dag}_{\mathbf{k}}(t)a_{\mathbf{k}}(t)\frac{dt}{d\tau} \nonumber\\
&&+\mu R_{2}^{A}(\tau)\phi(x_A(\tau))+\mu R_{2}^{B}(\tau)\phi(x_B(\tau))\;.
\label{Total-Hamiltonian}
\eea

To investigate the respective contributions of the vacuum fluctuations and the atomic radiation reaction to the interatomic potential, we must split the dynamical variables of the field and the atoms into free parts and source parts. As a first step, we should obtain the time dependence of the dynamical variables of the atoms and the field. By resorting to the total Hamiltonian of the ``atoms+field" system, Eq.~(\ref{Total-Hamiltonian}), we can get the Heisenberg equations of motion of the dynamical variables of atom $\xi$,
\bea
\frac{d}{d\tau}R_{\pm}^{\xi}(\tau)&=&\pm i\omega_{\xi}R_{\pm}^{\xi}(\tau)+i\mu \phi(x_{\xi}(\tau))[R_{2}^{\xi}(\tau),R_{\pm}^{\xi}(\tau)]\;,\\
\frac{d}{d\tau}R_{3}^{\xi}(\tau)&=&i\mu \phi(x_{\xi}(\tau))[R_{2}^{\xi}(\tau),R_{3}^{\xi}(\tau)]\;,
\eea
and that of the scalar field,
\beq
\frac{d}{d\tau}a_{\mathbf{k}}(t(\tau))=-i\omega_{\mathbf{k}}a_{\mathbf{k}}(t(\tau))\frac{dt}{d\tau}+i\mu \sum_{{\xi}=A}^{B}R_{2}^{\xi}(\tau)[\phi(x_{\xi}(\tau)),a_{\mathbf{k}}(t(\tau))]\;,
\eeq
where $[,]$ denotes the commutator of two operators.

Solving the above equations and splitting every solution into the free part  denoted by the superscript ``f"  which exists even when there is no coupling between the atoms and the field, and the source part denoted by the superscript ``s" which appears as a result of the atom-field interaction,  i.e.,
\bea
R_{\pm}^{\xi}(\tau)&=&R^{\xi,f}_{\pm}(\tau)+R^{\xi,s}_{\pm}(\tau)\;,
\label{R-pm}\\
R_{3}^{\xi}(\tau)&=&R^{\xi,f}_{3}(\tau)+R^{\xi,s}_{3}(\tau)\;,\\
a_{\mathbf{k}}(t(\tau))&=&a^f_{\mathbf{k}}(t(\tau))+a^s_{\mathbf{k}}(t(\tau))\;,
\eea
we have
\bea
R^{\xi,f}_{\pm}(\tau)=R^{\xi,f}_{\pm}(\tau_0)e^{\pm i\omega_{\xi}(\tau-\tau_0)}\;,\quad\quad\quad\quad\quad\quad\quad\quad\quad\quad\quad\label{free-sigma-pm}\\
R^{\xi,s}_{\pm}(\tau)=i\mu \int_{\tau_0}^{\tau}d\tau_1\phi(x_{\xi}(\tau_1))[R_{2}^{\xi}(\tau_1),R_{\pm}^{\xi}(\tau_1)]e^{\pm i\omega_{\xi}(\tau-\tau_1)}\;,\;
\label{source-sigma-pm}\\
R^{\xi,f}_{3}(\tau)=R^{\xi,f}_{3}(\tau_0)\;,\quad\quad\quad\quad\quad\quad\quad\quad\quad\quad\quad\quad\quad\quad\;\;\;\\
R^{\xi,s}_{3}(\tau)=i\mu\int_{\tau_0}^{\tau}d\tau_1\;[R_{2}^{\xi}(\tau_1),R_{3}^{\xi}(\tau_1)]\phi(x_{\xi}(\tau_1))\;,\quad\quad\quad\quad
\label{source-sigma-3}
\eea
and
\bea
a^f_{\mathbf{k}}(t(\tau))&=&a^f_{\mathbf{k}}(t(\tau_0))e^{-i\omega_{\mathbf{k}}(t(\tau)-t(\tau_0))}\;,\label{free-annihilation-operator}\\
a^s_{\mathbf{k}}(t(\tau))&=&i\mu\sum_{\xi=A}^B\int_{\tau_0}^{\tau}d\tau_1R_{2}^{\xi}(\tau_1)[\phi(x_{\xi}(\tau_1)),a_{\mathbf{k}}(t(\tau_1))]\nn\\
&&\times e^{-i\omega_{\mathbf{k}}(t(\tau)-t(\tau_1))}\;.
\label{source-annihilation-operator}
\eea
Here and after, $\tau_0$ denotes the onset time of the atom-field interaction. With the above $a^f_{\mathbf{k}}(t(\tau))$ and $a^s_{\mathbf{k}}(t(\tau))$, we can further split the field operator, $\phi(x(\tau))$, into the free field
\beq
\phi^f(x(\tau))=\int d^3{\mathbf{k}}\;g_{\mathbf{k}}[a^{f}_{\mathbf{k}}(t(\tau))e^{i\mathbf{k}\cdot\mathbf{x}}+a^{\dag f}_{\mathbf{k}}(t(\tau))e^{-i\mathbf{k}\cdot\mathbf{x}}]\;,
\label{free-field}
\eeq
and the source field
\beq
\phi^s(x(\tau))=\int d^3{\mathbf{k}}\;g_{\mathbf{k}}[a^s_{\mathbf{k}}(t(\tau))e^{i\mathbf{k}\cdot\mathbf{x}}+a^{\dag s}_{\mathbf{k}}(t(\tau))e^{-i\mathbf{k}\cdot\mathbf{x}}]\;.
\label{source-field}
\eeq
The source parts of the operators of the atoms and the field, Eqs.~(\ref{source-sigma-pm}), (\ref{source-sigma-3}), (\ref{source-annihilation-operator}) and (\ref{source-field}) are accurate expressions, and we can further perturbatively expand them in terms of the coupling constant $\mu$. 
Then,  for the operators of the atoms, we have, up to the third order
\bea
R_2^{\xi}(\tau)&=&R_2^{\xi,(0)}(\tau)+R_2^{\xi,(1)}(\tau)+R_2^{\xi,(2)}(\tau)+R_2^{\xi,(3)}(\tau)+O(\mu^4)\;,\nn\\\\
R_3^{\xi}(\tau)&=&R_3^{\xi,(0)}(\tau)+R_3^{\xi,(1)}(\tau)+R_3^{\xi,(2)}(\tau)+R_3^{\xi,(3)}(\tau)+O(\mu^4)\;,\nn\\
\eea
and for the field operator,
\beq
\phi(x(\tau))=\phi^{(0)}(x(\tau))+\phi^{(1)}(x(\tau))+\phi^{(2)}(x(\tau))+\phi^{(3)}(x(\tau))+O(\mu^4)\;.
\eeq
The concrete zeroth to the third order expansion terms of these operators are listed out in Appendix.~\ref{Appendix-A}.

\section{Contributions of vacuum fluctuations and atomic radiation reaction to the variation rate of the Hamiltonian of the atoms} 
\label{variation-rate}

To distinguish the contributions of  the vacuum fluctuations and the atomic radiation reaction to the interatomic potential, we should consider the respective contributions of the free field and the source field. With this in mind, we first consider the time evolution of the Hamiltonian of one atom, atom $A$, for example, in terms of the contributions of the free field and the source field.

The Heisenberg equation of motion of the Hamiltonian of atom $A$, $H_A(\tau)=\omega_AR^A_3(\tau)$, is given by
\beq
\frac{d}{d\tau}H_A(\tau)=i\mu\omega_A[R_2^A(\tau),R_3^A(\tau)]\phi(x_A(\tau))\;.
\eeq
Dividing the field operator on the right of the above equation into a free part and a source part, $\phi(x_A(\tau))=\phi^f(x_A(\tau))+\phi^s(x_A(\tau))$, the variation rate of the Hamiltonian of atom $A$ is accordingly divided into two parts:
\beq
\frac{d}{d\tau}H_A(\tau)={\biggl(\frac{d}{d\tau}H_A(\tau)\biggr)}_{vf}+{\biggl(\frac{d}{d\tau}H_A(\tau)\biggr)}_{rr}\;,
\eeq
where ${\left(\frac{d}{d\tau}H_A(\tau)\right)}_{vf}$ and ${\left(\frac{d}{d\tau}H_A(\tau)\right)}_{rr}$ are related with the free field and the source field respectively, and thus represent their respective contributions. Though the field operator and the operators of the atoms commute, the free part and the source part of the field operator  no longer commute with the operators of the atoms. Thus for the contributions of the free field, we have
\bea
\biggl(\frac{d}{d\tau}H_A(\tau)\biggr)_{vf}&=&i\mu\omega_A(\lambda\phi^f(x_A(\tau))[R_2^A(\tau),R_3^A(\tau)]\nn\\&&
+(1-\lambda)[R_2^A(\tau),R_3^A(\tau)])\phi^f(x_A(\tau))\;,
\eea
and for the contributions of the source field,
\bea
\biggl(\frac{d}{d\tau}H_A(\tau)\biggr)_{rr}&=&i\mu\omega_A(\lambda\phi^s(x_A(\tau))[R_2^A(\tau),R_3^A(\tau)]\nn\\&&
+(1-\lambda)[R_2^A(\tau),R_3^A(\tau)]\phi^s(x_A(\tau))\;,
\eea
where $\lambda$ is an arbitrary number characterizing the ambiguity in the ordering of the operators of the field and the atoms. Notice that though different values of the parameter $\lambda$ result in the same total variation rate, $\frac{d}{d\tau}H_A(\tau)$, it does lead to different contributions of the free field and the source field. Similar situations also exist for a small system coupled to a reservoir, for which Daliabard, Dupont-Roc and Cohen Tannoudji demonstrated that only when $\lambda={1\/2}$ can the operators of the contributions of the free field and the source field be Hermitian and thus possess independent physical meanings~\cite{DDC82,DDC84}. This operator ordering is the so called ``symmetric operator ordering".

 By adopting the symmetric ordering for the operators of the atoms and the field as suggested by Daliabard, Dupont-Roc and Cohen Tannoudji~\cite{DDC82,DDC84} , the contributions of the free field and the source field to the variation rate of the Hamiltonian of atom $A$ are reduced to
\bea
{\left(\frac{d}{d\tau}H_A(\tau)\right)}_{vf}&=&\frac{1}{2}i\mu\omega_A\{\phi^{f}(x_A(\tau)),[R_{2}^{A}(\tau),R_{3}^{A}(\tau)]\}\;,
\label{vf-rate-contribtuion-operator}\\
{\left(\frac{d}{d\tau}H_A(\tau)\right)}_{rr}&=&\frac{1}{2}i\mu\omega_A\{\phi^{s}(x_A(\tau)),[R_{2}^{A}(\tau),R_{3}^{A}(\tau)]\}\;,
\label{rr-rate-contribtuion-operator}
\eea
where $\{,\}$ denotes the anti-commutator of two operators.

By perturbatively expanding the operators of the atoms and the field in the above two equations in terms of the  coupling constant, we can further express ${\left(\frac{d}{d\tau}H_A(\tau)\right)}_{vf}$ and ${\left(\frac{d}{d\tau}H_A(\tau)\right)}_{rr}$ as
\beq
{\left(\frac{d}{d\tau}H_A(\tau)\right)}_{vf}=\sum_{j=1}^{\infty}{\left(\frac{d}{d\tau}H_A(\tau)\right)}_{vf}^{(j)}
\eeq
with the terms up to the fourth-order in the coupling constant given by
\beq
{\left(\frac{d}{d\tau}H_A(\tau)\right)}_{vf}^{(1)}=\frac{1}{2}i\mu\omega_A\{\phi^{f}(x_A(\tau)),[R_{2}^{A,f}(\tau),R_{3}^{A,f}(\tau)]\}\;,
\label{first-order-vf-Hamiltonian}
\eeq
\bea
{\left(\frac{d}{d\tau}H_A(\tau)\right)}_{vf}^{(2)}=\frac{1}{2}i\mu\omega_A\{\phi^{f}(x_A(\tau)),[R_{2}^{A,(1)}(\tau),R_{3}^{A,f}(\tau)]\}\nn\\
\quad\quad\quad\quad\quad\;\;+\frac{1}{2}i\mu\omega_A\{\phi^{f}(x_A(\tau)),[R_{2}^{A,f}(\tau),R_{3}^{A,(1)}(\tau)]\}\;,
\label{vf-contribution-second-order-operator}
\eea
\bea
{\left(\frac{d}{d\tau}H_A(\tau)\right)}_{vf}^{(3)}=\frac{1}{2}i\mu\omega_A\{\phi^{f}(x_A(\tau)),[R_{2}^{A,(1)}(\tau),R_{3}^{A,(1)}(\tau)]\}\nn\\
\quad\quad\quad\quad\quad\quad+\frac{1}{2}i\mu\omega_A\{\phi^{f}(x_A(\tau)),[R_{2}^{A,f}(\tau),R_{3}^{A,(2)}(\tau)]\}\nn\\
\quad\quad\quad\quad\quad\quad+\frac{1}{2}i\mu\omega_A\{\phi^{f}(x_A(\tau)),[R_{2}^{A,(2)}(\tau),R_{3}^{A,f}(\tau)]\}\;,
\eea
\bea
{\left(\frac{d}{d\tau}H_A(\tau)\right)}_{vf}^{(4)}=\frac{1}{2}i\mu\omega_A\{\phi^{f}(x_A(\tau)),[R_{2}^{A,f}(\tau),R_{3}^{A,(3)}(\tau)]\}\nn\\
\quad\quad\quad\quad\quad\quad+\frac{1}{2}i\mu\omega_A\{\phi^{f}(x_A(\tau)),[R_{2}^{A,(1)}(\tau),R_{3}^{A,(2)}(\tau)]\}\nn\\
\quad\quad\quad\quad\quad\quad+\frac{1}{2}i\mu\omega_A\{\phi^{f}(x_A(\tau)),[R_{2}^{A,(2)}(\tau),R_{3}^{A,(1)}(\tau)]\}\nn\\
\quad\quad\quad\quad\quad\quad+\frac{1}{2}i\mu\omega_A\{\phi^{f}(x_A(\tau)),[R_{2}^{A,(3)}(\tau),R_{3}^{A,f}(\tau)]\}\;,
\eea
and
\bea
{\left(\frac{d}{d\tau}H_A(\tau)\right)}_{rr}=\sum_{j=2}^{\infty}{\left(\frac{d}{d\tau}H_A(\tau)\right)}_{rr}^{(j)}
\eea
with
\beq
{\left(\frac{d}{d\tau}H_A(\tau)\right)}_{rr}^{(2)}=\frac{1}{2}i\mu\omega_A\{\phi^{(1)}(x_A(\tau)),[R_{2}^{A,f}(\tau),R_{3}^{A,f}(\tau)]\}\;,
\label{second-order-rr}
\eeq
\bea
{\left(\frac{d}{d\tau}H_A(\tau)\right)}_{rr}^{(3)}=\frac{1}{2}i\mu\omega_A\{\phi^{(1)}(x_A(\tau)),[R_{2}^{A,(1)}(\tau),R_{3}^{A,f}(\tau)]\}\nn\\
\quad\quad\quad\quad\quad+\frac{1}{2}i\mu\omega_A\{\phi^{(1)}(x_A(\tau)),[R_{2}^{A,f}(\tau),R_{3}^{A,(1)}(\tau)]\}\nn\\
\quad\quad\quad\quad\quad+\frac{1}{2}i\mu\omega_A\{\phi^{(2)}(x_A(\tau)),[R_{2}^{A,f}(\tau),R_{3}^{A,f}(\tau)]\}\;,
\eea
\begin{widetext}
\bea
{\left(\frac{d}{d\tau}H_A(\tau)\right)}_{rr}^{(4)}&=&\frac{1}{2}i\mu\omega_A\{\phi^{(1)}(x_A(\tau)),[R_{2}^{A,(1)}(\tau),R_{3}^{A,(1)}(\tau)]\}
+\frac{1}{2}i\mu\omega_A\{\phi^{(1)}(x_A(\tau)),[R_{2}^{A,f}(\tau),R_{3}^{A,(2)}(\tau)]\}\nn\\&&
+\frac{1}{2}i\mu\omega_A\{\phi^{(1)}(x_A(\tau)),[R_{2}^{A,(2)}(\tau),R_{3}^{A,f}(\tau)]\}
+\frac{1}{2}i\mu\omega_A\{\phi^{(2)}(x_A(\tau)),[R_{2}^{A,f}(\tau),R_{3}^{A,(1)}(\tau)]\}\nn\\&&
+\frac{1}{2}i\mu\omega_A\{\phi^{(2)}(x_A(\tau)),[R_{2}^{A,(1)}(\tau),R_{3}^{A,f}(\tau)]\}
+\frac{1}{2}i\mu\omega_A\{\phi^{(3)}(x_A(\tau)),[R_{2}^{A,f}(\tau),R_{3}^{A,f}(\tau)]\}\;.
\label{fourth-order-rr-Hamiltonian}
\eea
As we will show later, the leading interatomic potential of two ground-state atoms is of the fourth order. So, we do not go to higher orders here.

With the use  in Eqs.~(\ref{first-order-vf-Hamiltonian})-(\ref{fourth-order-rr-Hamiltonian}) of the perturbative expansions  of the operators $R_2^A$, $R_3^A$ and $\phi(x(\tau))$, which are given in Appendix.~\ref{Appendix-A}, the above variation rates can be further expanded in terms of the free operators of the atoms and the field. Evaluating their averages over the vacuum state of the field, we  obtain,  up to the fourth order of the atom-field coupling, the contributions of the vacuum fluctuations and the atomic radiation reaction to the variation rate of the Hamiltonian of atom $A$.

For the first order, we find that the contributions of vacuum fluctuations $\langle0|{\left(\frac{d}{d\tau}H_A(\tau)\right)}_{vf}^{(1)}|0\rangle=0$ because ${\left(\frac{d}{d\tau}H_A(\tau)\right)}_{vf}^{(1)}$ is of the first power of the free field operator [see Eq.~(\ref{first-order-vf-Hamiltonian})]. At this order the atomic radiation reaction never plays any part.

For the second order, Eq.~(\ref{vf-contribution-second-order-operator}) shows that the contributions of the free field to the variation rate of the Hamiltonian of atom $A$ results from the interaction between the free field and  atom $A$ which involves $R_2^{A,(1)}$ and $R_3^{A,(1)}$ in the first order. As indicated in Eqs.~(\ref{R2A-(1)}) and (\ref{R3A-(1)}), these operators in the first order are induced by the interaction between the free field and atom $A$, and thus are independent of atom $B$.  So is  $\langle0|{\left(\frac{d}{d\tau}H_A(\tau)\right)}_{vf}^{(2)}|0\rangle$.  As a result, there is no contribution to the interatomic potential from the vacuum fluctuations in the second order. While for the contributions of the atomic radiation reaction, we obtain
\bea
\langle0|{\left(\frac{d}{d\tau}H_A(\tau)\right)}_{rr}^{(2)}|0\rangle
={1\/2}(i\mu)^2\omega_A\int^{\tau}_{\tau_0}d\tau_1\langle0|[\phi^f(x_B(\tau_1)),\phi^f(x_A(\tau))]|0\rangle
\{R_2^{B,f}(\tau_1),[R_2^{A,f}(\tau),R_3^{A,f}(\tau)]\}+h(\tau)\;,
\label{vaccum-average-Hamiltonian-rr-(2)-o}
\eea
where $h(\tau)$ represents a collection of terms which are independent of atom $B$ and thus it  does not  contribute to the interatomic interaction. This equation shows that different from the contributions of the vacuum fluctuations, the contributions of the atomic radiation reaction, to the second order, are dependent on both atoms, and thus may contribute to the interatomic interaction. The dependence of $\langle0|{\left(\frac{d}{d\tau}H_A(\tau)\right)}_{rr}^{(2)}|0\rangle$ on atom $B$ is physically understandable. As is indicated in Eq.~(\ref{second-order-rr}), the contributions of the atomic radiation reaction to the variation rate of the Hamiltonian of atom $A$, in the second order, originate from the interaction between atom $A$ and the first order field, $\phi^{(1)}(x_A)$, which is the radiative field induced by the free field of both atoms [see Eq.~(\ref{first-order-field-operator}) in Appendix.~\ref{Appendix-A}].

For the third order, $\langle0|{\left(\frac{d}{d\tau}H_A(\tau)\right)}_{vf,rr}^{(3)}|0\rangle=0$ due to the fact that the number of free field operators in ${\left(\frac{d}{d\tau}H_A(\tau)\right)}_{vf,rr}^{(3)}$ is odd.

For the fourth order, we have, for the contributions of the vacuum fluctuations,
\bea
\langle0|{\left(\frac{d}{d\tau}H_A(\tau)\right)}_{vf}^{(4)}|0\rangle&=&
{1\/2}(i\mu)^4\omega_A\int^{\tau}_{\tau_0}d\tau_1\int^{\tau_1}_{\tau_0}d\tau_2\int^{\tau_2}_{\tau_0}d\tau_3
\langle0|\{\phi^f(x_A(\tau)),\phi^f(x_B(\tau_3))\}[\phi^f(x_B(\tau_2)),\phi^f(x_A(\tau_1))]|0\rangle\nn\\&&\times
[R_2^{B,f}(\tau_3),R_2^{B,f}(\tau_2)][R_2^{A,f}(\tau_1),[R_2^{A,f}(\tau),R_3^{A,f}(\tau)]]+f(\tau)\;,
\label{vaccum-average-Hamiltonian-vf-(4)-o}
\eea
where $f(\tau)$ stands for a collection  of terms of the zeroth/first power in $R_2^{B,f}$. As these terms are rather tedious and more importantly they do not  contribute to the interatomic interaction as we will explain later, we abbreviate them by $f(\tau)$ here.  Similar abbreviations in the contributions of the atomic radiation reaction will also be used in the remaining part of this section. For the fourth order contributions of the atomic radiation reaction, we have
\bea
&&\langle0|{\left(\frac{d}{d\tau}H_A(\tau)\right)}_{rr}^{(4)}|0\rangle={1\/2}(i\mu)^4\omega_A\nn\\&&
\times\biggl(\int^{\tau}_{\tau_0}d\tau_1\int^{\tau}_{\tau_0}d\tau_2\int^{\tau_2}_{\tau_0}d\tau_3
\langle0|[\phi^f(x_B(\tau_1)),\phi^f(x_A(\tau))][\phi^f(x_B(\tau_3)),\phi^f(x_A(\tau_2))]|0\rangle
\{R_2^{B,f}(\tau_1),R_2^{B,f}(\tau_3)\}[R_2^{A,f}(\tau_2),[R_2^{A,f}(\tau),R_3^{A,f}(\tau)]]\nn\\&&+
\int^{\tau}_{\tau_0}d\tau_1\int^{\tau_1}_{\tau_0}d\tau_2\int^{\tau}_{\tau_0}d\tau_3
\langle0|\{\phi^f(x_B(\tau_2)),\phi^f(x_A(\tau_3))\}[\phi^f(x_B(\tau_1)),\phi^f(x_A(\tau))]|0\rangle
[R_2^{B,f}(\tau_2),R_2^{B,f}(\tau_1)][R_2^{A,f}(\tau_3),[R_2^{A,f}(\tau),R_3^{A,f}(\tau)]]\nn\\&&+
\int^{\tau}_{\tau_0}d\tau_1\int^{\tau_1}_{\tau_0}d\tau_2\int^{\tau_1}_{\tau_0}d\tau_3
\langle0|[\phi^f(x_B(\tau_2)),\phi^f(x_A(\tau_1))][\phi^f(x_B(\tau_3)),\phi^f(x_A(\tau))]|0\rangle
[R_2^{B,f}(\tau_2),R_2^{B,f}(\tau_3)]\{R_2^{A,f}(\tau_1),[R_2^{A,f}(\tau),R_3^{A,f}(\tau)]\}\nn\\&&+
\int^{\tau}_{\tau_0}d\tau_1\int^{\tau_1}_{\tau_0}d\tau_2\int^{\tau_2}_{\tau_0}d\tau_3
\langle0|[\phi^f(x_B(\tau_2)),\phi^f(x_A(\tau_1))][\phi^f(x_B(\tau_3)),\phi^f(x_A(\tau))]|0\rangle
[R_2^{B,f}(\tau_3),R_2^{B,f}(\tau_2)]\{R_2^{A,f}(\tau_1),[R_2^{A,f}(\tau),R_3^{A,f}(\tau)]\}\nn\\&&+
\int^{\tau}_{\tau_0}d\tau_1\int^{\tau_1}_{\tau_0}d\tau_2\int^{\tau_2}_{\tau_0}d\tau_3
\langle0|[\phi^f(x_A(\tau_1)),\phi^f(x_B(\tau_3))][\phi^f(x_B(\tau_2)),\phi^f(x_A(\tau))]|0\rangle
[R_2^{B,f}(\tau_3),R_2^{B,f}(\tau_2)\}\{R_2^{A,f}(\tau_1),[R_2^{A,f}(\tau),R_3^{A,f}(\tau)]\}\nn\\&&+
\int^{\tau}_{\tau_0}d\tau_1\int^{\tau_1}_{\tau_0}d\tau_2\int^{\tau_2}_{\tau_0}d\tau_3
\langle0|[\phi^f(x_A(\tau_3)),\phi^f(x_B(\tau_2))][\phi^f(x_B(\tau_1)),\phi^f(x_A(\tau))]|0\rangle
[R_2^{B,f}(\tau_2),R_2^{B,f}(\tau_1)]\{R_2^{A,f}(\tau_3),[R_2^{A,f}(\tau),R_3^{A,f}(\tau)]\}\biggr)\nn\\&&
+g(\tau)\;,
\label{vaccum-average-Hamiltonian-rr-(4)-o}
\eea
\end{widetext}
where $g(\tau)$ is an abbreviation of a collection of terms of the zeroth/first/third power in $R_2^{B,f}$, and similar to  $f(\tau)$ in Eq.~(\ref{vaccum-average-Hamiltonian-vf-(4)-o}), it does not contribute to the interatomic potential.

Simplifying the average of the product of the commutator and the anticommutator of the field operators [see Eq.~(\ref{vaccum-average-Hamiltonian-vf-(4)-o})] over the vacuum state into the product of two two-point functions of the field,
\bea
\langle0|\{\phi^f(x_A(\tau)),\phi^f(x_B(\tau_3))\}[\phi^f(x_B(\tau_2)),\phi^f(x_A(\tau_1))]|0\rangle\nn\\
=4C^F(x_A(\tau),x_B(\tau_3))\;\chi^F(x_B(\tau_2),x_A(\tau_1))\quad\quad\quad
\eea
with $C^F(x_A(\tau),x_B(\tau_3))$ and $\chi^F(x_B(\tau_2),x_A(\tau_1))$ respectively being the symmetric correlation function and the linear susceptibility of the field defined by
\bea
C^F(x_{\xi}(\tau_j),x_{\xi'}(\tau_l))={1\/2}\langle0|\{\phi^f(x_{\xi}(\tau_j)),\phi^f(x_{\xi'}(\tau_l))\}|0\rangle\;,
\label{C-F}\\
\chi^F(x_{\xi}(\tau_j),x_{\xi'}(\tau_l))={1\/2}\langle0|[\phi^f(x_{\xi}(\tau_j)),\phi^f(x_{\xi'}(\tau_l))]|0\rangle\;,
\label{chi-F}
\eea
where $\xi,\xi'=A$ or $B$ and $j,l=1,2,3$, we can transform Eq.~(\ref{vaccum-average-Hamiltonian-vf-(4)-o}) into
\begin{widetext}
\bea
\langle0|{\left(\frac{d}{d\tau}H_A(\tau)\right)}_{vf}^{(4)}|0\rangle&=&
2\mu^4\omega_A\int^{\tau}_{\tau_0}d\tau_1\int^{\tau_1}_{\tau_0}d\tau_2\int^{\tau_2}_{\tau_0}d\tau_3\;
C^F(x_A(\tau),x_B(\tau_3))\;\chi^F(x_B(\tau_2),x_A(\tau_1))[R_2^{B,f}(\tau_3),R_2^{B,f}(\tau_2)]\nn\\&&\times
[R_2^{A,f}(\tau_1),[R_2^{A,f}(\tau),R_3^{A,f}(\tau)]]+f(\tau)\;,
\label{vaccum-average-Hamiltonian-vf-(4)}
\eea
which is the contributions of the vacuum fluctuations to the variation rate of the Hamiltonian of atom $A$. Here the variation  of the Hamiltonian of atom $A$ in the fourth order can be physically understood as follows. The interaction between the free field and atom $B$ induces a radiative field in the first order, and meanwhile, atom $B$ is endowed with an instantaneous  monopole $R_2^B(\tau_3)$ in the first order; then the first order radiative field acts back on atom $B$ and as a result induces a radiative field in the second order and a monopole in the second order, i.e.,  $R_2^{B}(\tau_2)$,  in atom $B$.
Next the action of this second order radiative field on atom $A$ produces a monopole  in the third order,  i.e., $R^A_2(\tau_1)$,  in atom $A$. Finally, the interaction between atom $A$ with the third order monopole  and the free field results in the above fourth order variation rate.

Similarly, the contributions of the atomic radiation reaction to the variation rate of the Hamiltonian of atom $A$ in the fourth order [Eq.~(\ref{vaccum-average-Hamiltonian-rr-(4)-o})], can be simplified to be:
\bea
&&\langle0|{\left(\frac{d}{d\tau}H_A(\tau)\right)}_{rr}^{(4)}|0\rangle\nn\\&=&
2\mu^4\omega_A\biggl(\int^{\tau}_{\tau_0}d\tau_1\int^{\tau}_{\tau_0}d\tau_2\int^{\tau_2}_{\tau_0}d\tau_3\;
\chi^F(x_B(\tau_1),x_A(\tau))\chi^F(x_B(\tau_3),x_A(\tau_2))
\{R_2^{B,f}(\tau_1),R_2^{B,f}(\tau_3)\}[R_2^{A,f}(\tau_2),[R_2^{A,f}(\tau),R_3^{A,f}(\tau)]]\nn\\&&+
\int^{\tau}_{\tau_0}d\tau_1\int^{\tau_1}_{\tau_0}d\tau_2\int^{\tau}_{\tau_0}d\tau_3\;
C^F(x_B(\tau_2),x_A(\tau_3))\chi^F(x_B(\tau_1),x_A(\tau))
[R_2^{B,f}(\tau_2),R_2^{B,f}(\tau_1)][R_2^{A,f}(\tau_3),[R_2^{A,f}(\tau),R_3^{A,f}(\tau)]]\nn\\&&+
\int^{\tau}_{\tau_0}d\tau_1\int^{\tau_1}_{\tau_0}d\tau_2\int^{\tau_1}_{\tau_0}d\tau_3\;
\chi^F(x_B(\tau_2),x_A(\tau_1))\chi^F(x_B(\tau_3),x_A(\tau))
[R_2^{B,f}(\tau_2),R_2^{B,f}(\tau_3)]\{R_2^{A,f}(\tau_1),[R_2^{A,f}(\tau),R_3^{A,f}(\tau)]\}\nn\\&&+
\int^{\tau}_{\tau_0}d\tau_1\int^{\tau_1}_{\tau_0}d\tau_2\int^{\tau_2}_{\tau_0}d\tau_3\;
\chi^F(x_B(\tau_2),x_A(\tau_1))\chi^F(x_B(\tau_3),x_A(\tau))
[R_2^{B,f}(\tau_3),R_2^{B,f}(\tau_2)]\{R_2^{A,f}(\tau_1),[R_2^{A,f}(\tau),R_3^{A,f}(\tau)]\}\nn\\&&+
\int^{\tau}_{\tau_0}d\tau_1\int^{\tau_1}_{\tau_0}d\tau_2\int^{\tau_2}_{\tau_0}d\tau_3\;
\chi^F(x_A(\tau_1),x_B(\tau_3))\chi^F(x_B(\tau_2),x_A(\tau))
[R_2^{B,f}(\tau_3),R_2^{B,f}(\tau_2)\}\{R_2^{A,f}(\tau_1),[R_2^{A,f}(\tau),R_3^{A,f}(\tau)]\}\nn\\&&+
\int^{\tau}_{\tau_0}d\tau_1\int^{\tau_1}_{\tau_0}d\tau_2\int^{\tau_2}_{\tau_0}d\tau_3\;
\chi^F(x_A(\tau_3),x_B(\tau_2))\chi^F(x_B(\tau_1),x_A(\tau))
[R_2^{B,f}(\tau_2),R_2^{B,f}(\tau_1)]\{R_2^{A,f}(\tau_3),[R_2^{A,f}(\tau),R_3^{A,f}(\tau)]\}\biggr)\nn\\&&
+g(\tau)\;.
\label{vaccum-average-Hamiltonian-rr-(4)}
\eea
\end{widetext}
Obviously, the above  contributions of the atomic radiation reaction to the variation rate of the Hamiltonian of atom $A$ are much more complex than the contributions of vacuum fluctuations. As can be seen from Eq.~(\ref{fourth-order-rr-Hamiltonian}), these contributions are  caused by the action of the first to third order radiative fields of the atoms on atom $A$. The physics underlying each term can also be understood similarly following what is done for Eq.~(\ref{vaccum-average-Hamiltonian-vf-(4)}).

The contributions of the vacuum fluctuations and the atomic radiation reaction to the variation rate of the Hamiltonian of atom $B$, up to the fourth order, can be easily obtained by exchanging  $A\rightleftharpoons B$ in Eqs.~(\ref{vaccum-average-Hamiltonian-rr-(2)-o}), (\ref{vaccum-average-Hamiltonian-vf-(4)}) and (\ref{vaccum-average-Hamiltonian-rr-(4)}).

\section{Effective Hamiltonians of the contributions of vacuum fluctuations and atomic radiation reaction to the interatomic interaction}
\label{eff-Hamiltonian-vf-rr}
To find the interatomic potential, we should next write the variation rate of the Hamiltonian of the two-atom system into the following form:
\bea
\langle0|{\left(\frac{d}{d\tau}H_S(\tau)\right)}_{vf,rr}|0\rangle&=&i[(H_S(\tau))^{eff}_{vf,rr},H_S^f(\tau)]\nn\\&&
+non-Hermitian\;terms\;,
\label{full-variation-rate}
\eea
where $H_S(\tau)$ is the Hamiltonian of the two-atom system defined in Eq.~(\ref{H_S}), and  $(H_S(\tau))^{eff}_{vf,rr}$ are the effective Hamiltonians of the contributions of vacuum fluctuations and atomic radiation reaction to the interatomic interaction we are searching for, which are dependent on both atoms. As the contributions of the vacuum fluctuations and the atomic radiation reaction to the variation rate of Hamiltonians of the atoms in the first order vanishes, we start our discussion with the second order.

\subsection{Second order effective Hamiltonians.}
\label{subsec-second-order}

For the second order, since the vacuum fluctuations do not contribute to the interatomic potential as we have explained in the paragraph above  Eq.~(\ref{vaccum-average-Hamiltonian-rr-(2)-o}), we turn our attention to the contributions of the atomic radiation reaction.

Adding up the contributions of the atomic radiation reaction of both atoms, $\langle0|{\left(\frac{d}{d\tau}H_A(\tau)\right)}_{rr}^{(2)}|0\rangle$ [see Eq.~(\ref{vaccum-average-Hamiltonian-rr-(2)-o})] and $\langle0|{\left(\frac{d}{d\tau}H_B(\tau)\right)}_{rr}^{(2)}|0\rangle$, we obtain the contributions of the atomic radiation reaction to the variation rate of the Hamiltonian of the two-atom system,
\begin{widetext}
\bea
\langle0|{\left(\frac{d}{d\tau}H_S(\tau)\right)}_{rr}^{(2)}|0\rangle
&=&(i\mu)^2\int^{\tau}_{\tau_0}d\tau_1\;\chi^F(x_B(\tau_1),x_A(\tau))\{R_2^{B,f}(\tau_1),[R_2^{A,f}(\tau),H_A^{f}(\tau)]\}\nn\\&&
+(i\mu)^2\int^{\tau}_{\tau_0}d\tau_1\;\chi^F(x_A(\tau_1),x_B(\tau))\{R_2^{A,f}(\tau_1),[R_2^{B,f}(\tau),H_B^{f}(\tau)]\}\;,
\eea
with $H_{\xi}^f(\tau)=\omega_{\xi}R_3^{\xi,f}(\tau)$. Hereafter,  we omit the terms in the variation rates of the Hamiltonians that are only dependent on one atom since we are concerned with the interatomic interaction. As $[R_2^{A,f}(\tau),H_A^{f}(\tau)]=[R_2^{A,f}(\tau),H_S^{f}(\tau)]$ and $[R_2^{B,f}(\tau),H_B^{f}(\tau)]=[R_2^{B,f}(\tau),H_S^{f}(\tau)]$, this equation can also be expressed as
\bea
\langle0|{\left(\frac{d}{d\tau}H_S(\tau)\right)}_{rr}^{(2)}|0\rangle
&=&(i\mu)^2\int^{\tau}_{\tau_0}d\tau_1\;\chi^F(x_B(\tau_1),x_A(\tau))\{R_2^{B,f}(\tau_1),[R_2^{A,f}(\tau),H_S^{f}(\tau)]\}\nn\\&&
+(i\mu)^2\int^{\tau}_{\tau_0}d\tau_1\;\chi^F(x_A(\tau_1),x_B(\tau))\{R_2^{A,f}(\tau_1),[R_2^{B,f}(\tau),H_S^{f}(\tau)]\}\;.
\label{H_S-rr-variation-rate-(2)}
\eea
To write the above equation in the form of Eq.~(\ref{full-variation-rate}) so as to read out the effective Hamiltonian of the contributions of the radiation reaction, we will appeal to the following relation,
\beq
\{O_1,[O_2,O_3]\}={1\/2}[\{O_1,O_2\},O_3]+O_2O_3O_1-O_1O_3O_2\label{O-2}\;,
\eeq
where $O_j$ ($j=1,2,3$) are three arbitrary operators satisfying  $[O_1,O_2]=0$, $[O_1,O_3]\neq0$ and $[O_2,O_3]\neq0$. An application of this relation to Eq.~(\ref{H_S-rr-variation-rate-(2)}) leads to
\bea
\langle0|{\left(\frac{d}{d\tau}H_S(\tau)\right)}_{rr}^{(2)}|0\rangle
&=&{1\/2}(i\mu)^2\int^{\tau}_{\tau_0}d\tau_1\;\chi^F(x_B(\tau_1),x_A(\tau)))[\{R_2^{B,f}(\tau_1),R_2^{A,f}(\tau)\},H_S^{f}(\tau)]\nn\\&&
+(i\mu)^2\int^{\tau}_{\tau_0}d\tau_1\;\chi^F(x_B(\tau_1),x_A(\tau)))
\biggl(R_2^{A,f}(\tau)H_S^{f}(\tau)R_2^{B,f}(\tau_1)-R_2^{B,f}(\tau_1)H_S^{f}(\tau)R_2^{A,f}(\tau)\biggl)\nn\\&&
+A\rightleftharpoons B\;terms\;.
\eea
Comparing it with Eq.~(\ref{full-variation-rate}), we conclude that the effective Hamiltonian of the contributions of the atomic radiation reaction to the interatomic interaction in the second order is
\bea
(H_S(\tau))_{rr}^{eff,(2)}&=&{1\/2}i\mu^2\int^{\tau}_{\tau_0}d\tau_1\;\chi^F(x_B(\tau_1),x_A(\tau))
\{R_2^{B,f}(\tau_1),R_2^{A,f}(\tau)\}
+A\rightleftharpoons B\;term\;.\quad
\label{effective-rr-(2)}
\eea
This effective Hamiltonian is composed of two terms with each term containing one $R^{A,f}_2$ and one $R_2^{B,f}$. To find the interatomic potential, we next should take the average of it over the state of the two-atom system, $|g_Ag_B\rangle$, and such an operation then gives rise to a vanishing result. So, $(H_S(\tau))_{rr}^{eff,(2)}$ does not contribute to the interatomic potential of two ground-state atoms. 
Thus, to calculate the interatomic potential of two ground-state atoms, we must go to higher orders.

Before going to higher orders directly, we now pause to give some comments on  our derivation of the above effective Hamiltonian. First, our approach here is  to write  the variation rate of the Hamiltonian of the two-atom system in the form of Eq.~(\ref{full-variation-rate}),  so that   the effective Hamiltonian concerned  is readily  read out from the commutator. This is realized by the use of the relation Eq.~(\ref{O-2}). However, during such a process, one should exercise  caution  and pay special attention to a tricky point we now explain. That is,  only when $[O_1,O_3]\neq0$ and $[O_2,O_3]\neq0$, can the first term on the right of Eq.~(\ref{O-2}) be the only commutator in the form of $[\cdots,O_3]$; otherwise, another commutator in this form may be hidden in the other terms. For example, $O_2O_3O_1-O_1O_3O_2$ on the right of Eq.~(\ref{O-2}),  can be written as ${1\/2}[\{O_1,O_2\},O_3]$ when $[O_1,O_3]=0$, $[O_1,O_2]=0$ and $[O_2,O_3]\neq0$, and in this case the effective Hamiltonian directly reads out from the commutator on the right of Eq.~(\ref{O-2}) would be incomplete.
Second, there is also an alternative approach to deriving the effective Hamiltonian  for the interatomic interaction,  that is, to  write  the variation rate of the Hamiltonian of each atom rather than that of the two-atom system  in the form of Eq.~(\ref{full-variation-rate}) and read out  a part of the effective Hamiltonian  for the interatomic interaction associated with the atom which we will denote by $(H_{\xi}(\tau))^{eff,(2)}_{rr}$ in the following discussion. Then one half of the summation, i.e., ${1\/2}((H_A(\tau))^{eff,(2)}_{rr}+(H_B(\tau))^{eff,(2)}_{rr})$, yields the effective Hamiltonian we are searching for. In so doing, one finds that
\bea
\langle0|{\left(\frac{d}{d\tau}H_A(\tau)\right)}_{rr}^{(2)}|0\rangle
&=&(i\mu)^2\int^{\tau}_{\tau_0}d\tau_1\;\chi^F(x_B(\tau_1),x_A(\tau))\{R_2^{B,f}(\tau_1),[R_2^{A,f}(\tau),H_A^{f}(\tau)]\}\nn\\
&=&(i\mu)^2\int^{\tau}_{\tau_0}d\tau_1\;\chi^F(x_B(\tau_1),x_A(\tau))[\{R_2^{B,f}(\tau_1),R_2^{A,f}(\tau)\},H_A^{f}(\tau)]\;.
\eea
Here we re-stress that care should be taken in rewriting the above equation  in the form of Eq.~(\ref{full-variation-rate}), by noting
\bea
\{R_2^{B,f}(\tau_1),[R_2^{A,f}(\tau),H_A^{f}(\tau)]\}&=&{1\/2}[\{R_2^{B,f}(\tau_1),R_2^{A,f}(\tau)\},H_A^{f}(\tau)]+R_2^{A,f}(\tau)H_A^{f}(\tau)R_2^{B,f}(\tau_1)
-R_2^{B,f}(\tau_1)H_A^{f}(\tau)R_2^{A,f}(\tau)\nn\\
&=&[\{R_2^{B,f}(\tau_1),R_2^{A,f}(\tau)\},H_A^{f}(\tau)]
\eea
\end{widetext}
as a result of $[R_2^{B,f}(\tau_1),R_2^{A,f}(\tau)]=0$, $[R_2^{A,f}(\tau),H_A^{f}(\tau)]\neq0$ and $[R_2^{B,f}(\tau_1),H_A^{f}(\tau)]=0$ [see our afore-discussed  example]. Now a comparison of this equation with
\bea
\langle0|{\left(\frac{d}{d\tau}H_A(\tau)\right)}_{rr}|0\rangle&=&i[(H_A(\tau))^{eff}_{rr},H_A^f(\tau)]\nn\\&&
+non-Hermitian\;terms\;,
\eea
leads to
\bea
(H_A(\tau))^{eff,(2)}_{rr}&=&i\mu^2\int^{\tau}_{\tau_0}d\tau_1\;\chi^F(x_B(\tau_1),x_A(\tau))\nn\\&&\times
\{R_2^{B,f}(\tau_1),R_2^{A,f}(\tau)\}\;.
\eea
Similarly, we can find out $(H_B(\tau))^{eff,(2)}_{rr}$. Then one half of the summation of $(H_A(\tau))^{eff,(2)}_{rr}$ and $(H_B(\tau))^{eff,(2)}_{rr}$ gives rise to an effective Hamiltonian  identical  to Eq.~(\ref{effective-rr-(2)}). So,  the two approaches in deriving the effective Hamiltonians are equivalent.  In the following section which deals with higher order effective Hamiltonians for the contributions of the vacuum fluctuations and the atomic radiation reaction, we adopt the first approach.

\subsection{Fourth order effective Hamiltonians.}

As we have shown in section.~\ref{variation-rate}, the vacuum fluctuations and the atomic radiation reaction, in the third order of the coupling constant, do not contribute to the variation rate of the Hamiltonian of the atoms.  So, we now turn our attention to their contributions  in the fourth order.
By using Eq.~(\ref{vaccum-average-Hamiltonian-vf-(4)}) and its counterpart of atom $B$, the contributions of the vacuum fluctuations to the variation rate of the Hamiltonian of the two-atom system can be written as
\begin{widetext}
\bea
\langle0|{\left(\frac{d}{d\tau}H_S(\tau)\right)}_{vf}^{(4)}|0\rangle
&=&\langle0|{\left(\frac{d}{d\tau}H_A(\tau)\right)}_{vf}^{(4)}|0\rangle+\langle0|{\left(\frac{d}{d\tau}H_B(\tau)\right)}_{vf}^{(4)}|0\rangle\nn\\
&=&2\mu^4\omega_A\int^{\tau}_{\tau_0}d\tau_1\int^{\tau_1}_{\tau_0}d\tau_2\int^{\tau_2}_{\tau_0}d\tau_3\;
C^F(x_A(\tau),x_B(\tau_3))\;\chi^F(x_B(\tau_2),x_A(\tau_1))[R_2^{B,f}(\tau_3),R_2^{B,f}(\tau_2)]\nn\\&&\times
[R_2^{A,f}(\tau_1),[R_2^{A,f}(\tau),R_3^{A,f}(\tau)]]+f(\tau)
+A\rightleftharpoons B\;terms\;.
\label{vf-(4)-two-atom-variation-rate}
\eea
Splitting the triple integral in the above equation into two equal parts, and using the Heisenberg equations of motion of $R_2^{A}$ and $R_2^B$ and doing the integrals by parts for one of them, we then get
\bea
&&\langle0|{\left(\frac{d}{d\tau}H_S(\tau)\right)}_{vf}^{(4)}|0\rangle\nn\\&=&
\mu^4\int^{\tau}_{\tau_0}d\tau_1\int^{\tau_1}_{\tau_0}d\tau_2\int^{\tau_2}_{\tau_0}d\tau_3\;
C^F(x_A(\tau),x_B(\tau_3))\;\chi^F(x_B(\tau_2),x_A(\tau_1))[R_2^{B,f}(\tau_3),R_2^{B,f}(\tau_2)][R_2^{A,f}(\tau_1),[R_2^{A,f}(\tau),H^f_A(\tau)]]\nn\\&&
+\mu^4\int^{\tau}_{\tau_0}d\tau_1\int^{\tau_1}_{\tau_0}d\tau_2\int^{\tau_2}_{\tau_0}d\tau_3\;
C^F(x_A(\tau),x_B(\tau_3))\;\chi^F(x_B(\tau_2),x_A(\tau_1))[R_2^{B,f}(\tau_3),[R_2^{B,f}(\tau_2),H^f_B(\tau)]][R_2^{A,f}(\tau_1),R_2^{A,f}(\tau)]\nn\\&&
+f(\tau)+A\rightleftharpoons B\;terms\;.\quad
\label{H_A-vf-variation-rate}
\eea
Next by resorting to the relation
\beq
[O_1,[O_2,O_3]]={1\/2}[[O_1,O_2],O_3]+{1\/2}\{\{O_1,O_2\},O_3\}-O_2O_3O_1-O_1O_3O_2\;,\label{O-1}
\eeq
where $O_j\;(j=1,2,3)$ are three arbitrary operators which do not commute with one another, we can further transform Eq.~(\ref{H_A-vf-variation-rate}) into
\bea
&&\langle0|{\left(\frac{d}{d\tau}H_S(\tau)\right)}_{vf}^{(4)}|0\rangle\nn\\&=&
{1\/2}\mu^4\int^{\tau}_{\tau_0}d\tau_1\int^{\tau_1}_{\tau_0}d\tau_2\int^{\tau_2}_{\tau_0}d\tau_3\;
C^F(x_A(\tau),x_B(\tau_3))\;\chi^F(x_B(\tau_2),x_A(\tau_1))[[R_2^{B,f}(\tau_3),R_2^{B,f}(\tau_2)][R_2^{A,f}(\tau_1),R_2^{A,f}(\tau)],H^f_S(\tau)]\nn\\&&
+\mu^4\int^{\tau}_{\tau_0}d\tau_1\int^{\tau_1}_{\tau_0}d\tau_2\int^{\tau_2}_{\tau_0}d\tau_3\;
C^F(x_A(\tau),x_B(\tau_3))\;\chi^F(x_B(\tau_2),x_A(\tau_1))[R_2^{B,f}(\tau_3),R_2^{B,f}(\tau_2)]\nn\\&&\quad\quad\quad\quad\quad\quad\times
\biggl({1\/2}\{\{R_2^{A,f}(\tau_1),R_2^{A,f}(\tau)\},H^f_A(\tau)\}-R_2^{A,f}(\tau)H^f_A(\tau)R_2^{A,f}(\tau_1)-R_2^{A,f}(\tau_1)H^f_A(\tau)R_2^{A,f}(\tau)\biggr)\nn\\&&
+\mu^4\int^{\tau}_{\tau_0}d\tau_1\int^{\tau_1}_{\tau_0}d\tau_2\int^{\tau_2}_{\tau_0}d\tau_3\;
C^F(x_A(\tau),x_B(\tau_3))\;\chi^F(x_B(\tau_2),x_A(\tau_1))[R_2^{A,f}(\tau_1),R_2^{A,f}(\tau)]\nn\\&&\quad\quad\quad\quad\quad\quad\times
\biggl({1\/2}\{\{R_2^{B,f}(\tau_3),R_2^{B,f}(\tau_2)\},H^f_B(\tau)\}-R_2^{B,f}(\tau_2)H^f_B(\tau)R_2^{B,f}(\tau_3)-R_2^{B,f}(\tau_3)H^f_B(\tau)R_2^{B,f}(\tau_2)\biggr)+{f}(\tau)\nn\\&&
+A\rightleftharpoons B\;terms
\label{vf-full-(4)}
\eea
Here, as  three operators $R_2^{A,f}(\tau_1),R_2^{A,f}(\tau)$, and $H^f_A(\tau)$ as well as   $R_2^{B,f}(\tau_3),R_2^{B,f}(\tau_2)$ and $H^f_B(\tau)$ do not commute with one another, the commutator in the second line of the above equation is the  only commutator on the right of Eq.~(\ref{vf-full-(4)}) in the form of $[\cdots,H^f_S(\tau)]$.
Comparing this equation with Eq.~(\ref{full-variation-rate}),
it is obvious that
\bea
(H_S(\tau))^{eff,(4)}_{vf}&=&{1\/2}i\mu^4\int^{\tau}_{\tau_0}d\tau_1\int^{\tau_1}_{\tau_0}d\tau_2\int^{\tau_2}_{\tau_0}d\tau_3\;
C^F(x_A(\tau),x_B(\tau_3))\;\chi^F(x_B(\tau_2),x_A(\tau_1))
[R_2^{B,f}(\tau_3),R_2^{B,f}(\tau_2)][R_2^{A,f}(\tau),R_2^{A,f}(\tau_1)]\nn\\&&
+A\rightleftharpoons B\;term\;,
\label{effective-Hamiltonian-vf}
\eea
which is the effective Hamiltonian of the contributions of the vacuum fluctuations to the interatomic interaction.

Similar operations on the contributions of the atomic radiation reaction to the variation rate of the Hamiltonian of the two atoms, give rise to the following effective Hamiltonian of the contributions of the atomic radiation reaction to the interatomic interaction
\bea
(H_S(\tau))_{rr}^{eff,(4)}&=&{1\/2}i\mu^4\int^{\tau}_{\tau_0}d\tau_1\int^{\tau}_{\tau_0}d\tau_2\int^{\tau_2}_{\tau_0}d\tau_3\;
\chi^F(x_B(\tau_1),x_A(\tau))\chi^F(x_B(\tau_3),x_A(\tau_2))
\{R_2^{B,f}(\tau_1),R_2^{B,f}(\tau_3)\}[R_2^{A,f}(\tau),R_2^{A,f}(\tau_2)]\nn\\&&
+{1\/2}i\mu^4\int^{\tau}_{\tau_0}d\tau_1\int^{\tau_1}_{\tau_0}d\tau_2\int^{\tau}_{\tau_0}d\tau_3\;
C^F(x_B(\tau_2),x_A(\tau_3))\chi^F(x_B(\tau_1),x_A(\tau))
[R_2^{B,f}(\tau_2),R_2^{B,f}(\tau_1)][R_2^{A,f}(\tau),R_2^{A,f}(\tau_3)]\nn\\&&
+{1\/2}i\mu^4\int^{\tau}_{\tau_0}d\tau_1\int^{\tau_1}_{\tau_0}d\tau_2\int^{\tau_1}_{\tau_0}d\tau_3\;
\chi^F(x_B(\tau_2),x_A(\tau_1))\chi^F(x_B(\tau_3),x_A(\tau))
[R_2^{B,f}(\tau_3),R_2^{B,f}(\tau_2)]\{R_2^{A,f}(\tau_1),R_2^{A,f}(\tau)\}\nn\\&&
+{1\/2}i\mu^4\int^{\tau}_{\tau_0}d\tau_1\int^{\tau_1}_{\tau_0}d\tau_2\int^{\tau_2}_{\tau_0}d\tau_3\;
\chi^F(x_B(\tau_2),x_A(\tau_1))\chi^F(x_B(\tau_3),x_A(\tau))
[R_2^{B,f}(\tau_2),R_2^{B,f}(\tau_3)]\{R_2^{A,f}(\tau_1),R_2^{A,f}(\tau)\}\nn\\&&
+{1\/2}i\mu^4\int^{\tau}_{\tau_0}d\tau_1\int^{\tau_1}_{\tau_0}d\tau_2\int^{\tau_2}_{\tau_0}d\tau_3\;
\chi^F(x_A(\tau_1),x_B(\tau_3))\chi^F(x_B(\tau_2),x_A(\tau))
[R_2^{B,f}(\tau_2),R_2^{B,f}(\tau_3)\}\{R_2^{A,f}(\tau_1),R_2^{A,f}(\tau)\}\nn\\&&
+{1\/2}i\mu^4\int^{\tau}_{\tau_0}d\tau_1\int^{\tau_1}_{\tau_0}d\tau_2\int^{\tau_2}_{\tau_0}d\tau_3\;
\chi^F(x_A(\tau_3),x_B(\tau_2))\chi^F(x_B(\tau_1),x_A(\tau))
[R_2^{B,f}(\tau_1),R_2^{B,f}(\tau_2)]\{R_2^{A,f}(\tau_3),R_2^{A,f}(\tau)\}\nn\\&&
+A\rightleftharpoons B\;terms\;.
\label{effective-rr}
\eea
Here on the right of the above equation we have also omitted terms which are of the zeroth/first/third power in $R_2^{B,f}$, as they are very tedious and do not contribute to the interatomic potential.

\section{Contributions of vacuum fluctuations and atomic radiation reaction to the interatomic potential of two ground-state atoms}
With the effective Hamiltonians of the contributions of the vacuum fluctuations and the atomic radiation reaction derived in the above section, we are now ready to evaluate the interatomic potential.
Taking the average of Eq.~(\ref{effective-Hamiltonian-vf}) over the state of the two-atom system, $|g_Ag_B\rangle$, we find that the contributions of the vacuum fluctuations to the interatomic potential is
\bea
(\delta E)_{vf}&=&2i\mu^4\int^{\tau}_{\tau_0}d\tau_1\int^{\tau_1}_{\tau_0}d\tau_2\int^{\tau_2}_{\tau_0}d\tau_3\;
C^F(x_A(\tau),x_B(\tau_3))\;\chi^F(x_B(\tau_2),x_A(\tau_1))\chi^B(\tau_3,\tau_2)\chi^A(\tau,\tau_1)\nn\\&&
+A\rightleftharpoons B\;term\;,
\label{vf-interatomic-potential}
\eea
where $\chi^{\xi}(\tau_j,\tau_l)$ is the antisymmetric statistical function of atom $\xi$ defined by
\bea
\chi^{\xi}(\tau_j,\tau_l)&=&\frac{1}{2}\langle g_{\xi}|[R^{\xi,f}_{2}(\tau_j),R_{2}^{\xi,f}(\tau_l)]|g_{\xi}\rangle\;.
\label{chi-atom-definnition}
\eea
Similarly, taking the average of Eq.~(\ref{effective-rr}) over the state of the two-atom system, $|g_Ag_B\rangle$, we get the contributions of the atomic radiation reaction to the interatomic potential:
\bea
(\delta E)_{rr}&=&2i\mu^4\int^{\tau}_{\tau_0}d\tau_1\int^{\tau}_{\tau_0}d\tau_2\int^{\tau_2}_{\tau_0}d\tau_3\;
\chi^F(x_B(\tau_1),x_A(\tau))\chi^F(x_B(\tau_3),x_A(\tau_2))C^B(\tau_1,\tau_3)\chi^A(\tau,\tau_2)\nn\\&&
+2i\mu^4\int^{\tau}_{\tau_0}d\tau_1\int^{\tau_1}_{\tau_0}d\tau_2\int^{\tau}_{\tau_0}d\tau_3\;
C^F(x_B(\tau_2),x_A(\tau_3))\chi^F(x_B(\tau_1),x_A(\tau))\chi^B(\tau_1,\tau_2))\chi^A(\tau_3,\tau)\nn\\&&
+2i\mu^4\int^{\tau}_{\tau_0}d\tau_1\int^{\tau_1}_{\tau_0}d\tau_2\int^{\tau_1}_{\tau_0}d\tau_3\;
\chi^F(x_B(\tau_2),x_A(\tau_1))\chi^F(x_B(\tau_3),x_A(\tau))\chi^B(\tau_3,\tau_2)C^A(\tau_1,\tau)\nn\\&&
+2i\mu^4\int^{\tau}_{\tau_0}d\tau_1\int^{\tau_1}_{\tau_0}d\tau_2\int^{\tau_2}_{\tau_0}d\tau_3\;
\chi^F(x_B(\tau_2),x_A(\tau_1))\chi^F(x_B(\tau_3),x_A(\tau))\chi^B(\tau_2,\tau_3)C^A(\tau_1,\tau)\nn\\&&
+2i\mu^4\int^{\tau}_{\tau_0}d\tau_1\int^{\tau_1}_{\tau_0}d\tau_2\int^{\tau_2}_{\tau_0}d\tau_3\;
\chi^F(x_A(\tau_1),x_B(\tau_3))\chi^F(x_B(\tau_2),x_A(\tau))\chi^B(\tau_2,\tau_3)C^A(\tau_1,\tau)\nn\\&&
+2i\mu^4\int^{\tau}_{\tau_0}d\tau_1\int^{\tau_1}_{\tau_0}d\tau_2\int^{\tau_2}_{\tau_0}d\tau_3\;
\chi^F(x_A(\tau_3),x_B(\tau_2))\chi^F(x_B(\tau_1),x_A(\tau))\chi^B(\tau_1,\tau_2)C^A(\tau_3,\tau)\nn\\&&
+A\rightleftharpoons B\;terms\;,
\label{rr-interatomic-potential}
\eea
\end{widetext}
where $C^{\xi}(\tau_j,\tau_l)$ is the symmetric statistical function of atom $\xi$ defined by
\beq
C^{\xi}(\tau_j,\tau_l)=\frac{1}{2}\langle g_{\xi}|\{R^{\xi,f}_{2}(\tau_j),R_{2}^{\xi,f}(\tau_l)\}|g_{\xi}\rangle\;.
\label{C-atom-definnition}
\eeq
Here we must point out that the contributions of the atomic radiation reaction to the interatomic potential of two ground-state atoms derived in this paper, Eq.~(\ref{rr-interatomic-potential}), is quite different from that in Ref.~\cite{Menezes17}, as there only one term which is identical to the last term in our Eq.~(\ref{rr-interatomic-potential}) exists while other five terms are absent [see Eq.~(4) in Ref.~\cite{Menezes17} where  the details of the derivation were not  given]. Probably this discrepancy may result from different expressions of the source parts of the dynamical variables of the atoms and the field [$R^{\xi,(i)}_2(\tau)$, $R^{\xi,(i)}_3(\tau)$ and $a^{(i)}_{\mathbf{k}}(\tau)$ ($\xi=A,B,\;i=1,2,3,\cdots$)].

Then the total interatomic potential can be obtained by summing up the contributions of the vacuum fluctuations and the atomic radiation reaction,
\beq
\delta E=(\delta E)_{vf}+(\delta E)_{rr}\;.
\eeq

For two static ground-state atoms in interaction with the fluctuating massless scalar field in vacuum, we have calculated, as an extra validity check, the interatomic potential with the fourth order DDC formalism derived in this paper, and we find that the result obtained is accurately the same as that obtained by using the perturbation theory. Now, it is worth pointing out that though we have assumed two nonidentical  atoms in our derivation, our fourth order DDC formalism is also applicable  to the case of two identical atoms. One can show that the use of the fourth order DDC formalism in the limit of equal transition frequency leads to an interatomic potential for two identical static ground-state atoms, the same  as that given by the perturbation theory. 

A few comments  are now in order. First, though  atoms in interaction with quantum scalar fields and those with  quantum electromagnetic fields which is more realistic describe distinctive situations in general, the scalar case can still model some special circumstance  of the  electromagnetic case.  For an example, when the two atoms are  polarizable  along the  same direction, then only the component  of the electromagnetic field along that direction plays a part in the interatomic interaction. This case can clearly be modeled by the scalar field model.

Second, let us point out that  a generalization of our result to the more realistic case with  vacuum electromagnetic fields is straightforward. Even before such a generalization is performed,  we can still  infer in general that some properties of the scalar field case are kept.
For example,  the interatomic interaction can  be ascribed to the vacuum field fluctuations and the atomic radiation reaction since the coupling between fields and the atom is linear in both the scalar and electromagnetic cases. Note however that if the atoms-field coupling is nonlinear, things become quite different.  For instance, when an atom is coupled to the Dirac field or the Rarita-Schwinger field, where even the simplest coupling is nonlinear,
the radiative properties can no longer be solely attributed to the field fluctuations and the atomic radiation reaction~\cite{Zhou12,Li14}. Of course, there are also features that are exclusively due to the EM. For example, in the electromagnetic field case,  the atomic dipole moments and the electromagnetic field are vectors as opposed to the monopole and the scalar filed which are scalars, and as a result, the interatomic potential is generally  dependent on the orientation of polarization of the atoms.

Finally, let us note that the fourth order DDC formalism treats the interatomic potential from a perspective different from that of the perturbation theory. The perturbation treatment calculates the interatomic potential by summing up the contributions of all the combinations of the intermediate states [the contributions of the exchange of photons in the interaction] without distinguishing the contributions of the vacuum fluctuations and the atomic radiation reaction. For example,  it has been shown that for two ground-state atoms there are twelve combinations of the intermediate states which contribute to the interatomic potential~\cite{Craig98}, and to derive this interatomic potential, one has to sum up all their contributions. While the DDC formalism calculates the contributions of the field fluctuations and atomic radiation reaction respectively without explicitly  referring   to the exchange of the photons in the interaction. By using the DDC formalism, the contributions of the vacuum fluctuations and the contributions of the atomic radiation reaction and thus their total effect, are finally expressed in terms of correlation functions of the field and statistical functions of the atoms.

\section{summary}
In this paper, we have given a detailed derivation of the fourth order DDC formalism for the interatomic potential of two ground-state atoms in interaction with the fluctuating scalar field in vacuum.

Starting from the Hamiltonian of the ``atoms+field" system, we first derive the Heisenberg equations of motion of the dynamical variables of the atoms and the field. After solving these equations, we split every solution into a free part which exists even when there is no coupling between the atoms and the field, and a source part which results from the interaction between the atoms and the field. In order to study the interatomic interaction order by order, we  perturbatively expand the operators of the atoms and the field in terms of the coupling constant. We adopt the symmetric operator ordering for the operators of the atoms and the field  to identify the respective contributions of the vacuum fluctuations and the atomic radiation reaction to the interatomic interaction.  By investigating the average variation rate of the Hamiltonian of the two-atom system in terms of the contributions of the free field and the source field, we derive the effective Hamiltonians of the contributions of the vacuum fluctuations and the contributions of the atomic radiation reaction. Finally,  we obtain, by taking the averages of the effective Hamiltonians over the state of the two-atom system,  the general formulae for the contributions of the vacuum fluctuations and the atomic radiation reaction to the interatomic potential, which can serve as the basis for the future research using DDC formalism  on the fourth-order effects in particular circumstances.

\begin{acknowledgments}
This work was supported in part by the NSFC under Grants
No. 11690034, No. 11875172 , No. 11405091 and No. 12075084;
and the K. C. Wong Magna Fund in Ningbo University.
\end{acknowledgments}

\appendix
\section{The operators of the atoms and the field in series.}
\label{Appendix-A}
In Eqs.~(\ref{source-sigma-pm}), (\ref{source-sigma-3}) and (\ref{source-field}), iterating the field operators and the operators of the atoms, $R_2^{\xi}$ and $R_3^{\xi}$, into their free parts and source parts and using the relation Eq.~(\ref{sigma-2}), we can expand the field operator and the operators of the atoms in terms of the coupling constant.

For $R_2^A(\tau)$, we have
\beq
R_2^A(\tau)=R_2^{A,(0)}(\tau)+R_2^{A,(1)}(\tau)+R_2^{A,(2)}(\tau)+R_2^{A,(3)}(\tau)+O(\mu^4)
\eeq
with
\bea
R_2^{A,(0)}(\tau)&=&R_2^{A,f}(\tau)\nn\\
&=&{i\/2}\left(R_-^{A,f}(\tau_0)e^{-i\omega_A(\tau-\tau_0)}-R_+^{A,f}(\tau_0)e^{i\omega_A(\tau-\tau_0)}\right),\quad\\
R_2^{A,(1)}(\tau)&=&i\mu\int^{\tau}_{\tau_0}d\tau_1\phi^f(x_A(\tau_1))[R_2^{A,f}(\tau_1),R_2^{A,f}(\tau)]\;,\label{R2A-(1)}
\eea
\begin{widetext}
\bea
R^{A,(2)}_2(\tau)&=&{1\/2}(i\mu)^2\int^{\tau}_{\tau_0}d\tau_1\int^{\tau_1}_{\tau_0}d\tau_2[\phi^f(x_A(\tau_2)),\phi^f(x_A(\tau_1))]
\{R_2^{A,f}(\tau_2),[R_2^{A,f}(\tau_1),R_2^{A,f}(\tau)]\}\nn\\
&&+(i\mu)^2\int^{\tau}_{\tau_0}d\tau_1\int^{\tau_1}_{\tau_0}d\tau_2[\phi^f(x_B(\tau_2)),\phi^f(x_A(\tau_1))]
R_2^{B,f}(\tau_2)[R_2^{A,f}(\tau_1),R_2^{A,f}(\tau)]\nn\\
&&+{1\/2}(i\mu)^2\int^{\tau}_{\tau_0}d\tau_1\int^{\tau_1}_{\tau_0}d\tau_2\{\phi^f(x_A(\tau_2)),\phi^f(x_A(\tau_1))\}
[R_2^{A,f}(\tau_2),[R_2^{A,f}(\tau_1),R_2^{A,f}(\tau)]]
\eea
and
\bea
R^{A,(3)}_2(\tau)&=&(i\mu)^3\int^{\tau}_{\tau_0}d\tau_1\int^{\tau_1}_{\tau_0}d\tau_2\int^{\tau_2}_{\tau_0}d\tau_3\phi^f(x_B(\tau_3))
[\phi^f(x_B(\tau_2)),\phi^f(x_A(\tau_1))][R_2^{B,f}(\tau_2),R_2^{B,f}(\tau_1)][R_2^{A,f}(\tau_1),R_2^{A,f}(\tau)]\nn\\&&
+\;terms\;of\;the\;zeroth/first\;power\;of\;R_2^{B,f}\;\;.
\eea
\end{widetext}
Here for  $R^{A,(3)}_2(\tau)$, we do not give the full expression but only list out the term of the second power in $R_2^{B,f}$, as the full expression is too lengthy,  and the other terms  are of the zeroth/first power in $R_2^{B,f}$  and do not contribute to the interatomic potential.

Similarly, for $R^{A}_3(\tau)$, we have:
\beq
R_{3}^{A}(\tau)=R^{A,(0)}_{3}(\tau)+R^{A,(1)}_{3}(\tau)+R^{A,(2)}_{3}(\tau)+R^{A,(3)}_{3}(\tau)+O(\mu^4)
\eeq
with
\bea
R^{A,(0)}_{3}(\tau)=R^{A,f}_{3}(\tau)\;,\quad\quad\quad\quad\quad\quad\quad\quad\quad\quad\quad\quad\;\\
R_3^{A,(1)}(\tau)=i\mu\int^{\tau}_{\tau_0}d\tau_1\phi^f(x_A(\tau_1))[R_2^{A,f}(\tau_1),R_3^{A,f}(\tau)]\;,\label{R3A-(1)}
\eea
\begin{widetext}
\bea
R^{A,(2)}_{3}(\tau)&=&\frac{1}{2}(i\mu)^2\int_{\tau_0}^{\tau}d\tau_1\int_{\tau_0}^{\tau_1}d\tau_2
[\phi^{f}(x_A(\tau_2)),\phi^{f}(x_A(\tau_1))]\{R^{A,f}_{2}(\tau_2),[R^{A,f}_{2}(\tau_1),R^{A,f}_{3}(\tau)]\}\nonumber\\
&&+(i\mu)^2\int_{\tau_0}^{\tau}d\tau_1\int_{\tau_0}^{\tau_1}d\tau_2
[\phi^{f}(x_B(\tau_2)),\phi^{f}(x_A(\tau_1))]R^{B,f}_{2}(\tau_2)[R^{A,f}_{2}(\tau_1),R^{A,f}_{3}(\tau)]\nn\\&&
+\frac{1}{2}(i\mu)^2\int_{\tau_0}^{\tau}d\tau_1\int_{\tau_0}^{\tau_1}d\tau_2
\{\phi^{f}(x_A(\tau_1)),\phi^{f}(x_A(\tau_2))\}[R^{A,f}_{2}(\tau_2),[R^{A,f}_{2}(\tau_1),R^{A,f}_{3}(\tau)]]\;,
\eea
and
\bea
R^{A,(3)}_3(\tau)&=&(i\mu)^3\int^{\tau}_{\tau_0}d\tau_1\int^{\tau_1}_{\tau_0}d\tau_2\int^{\tau_2}_{\tau_0}d\tau_3
\phi^f(x_B(\tau_3))[\phi^f(x_B(\tau_2)),\phi^f(x_A(\tau_1))][R_2^{B,f}(\tau_3),R_2^{B,f}(\tau_2)][R_2^{A,f}(\tau_1),R_3^{A,f}(\tau)]\nn\\
&&+\;terms\;of\;the\;zeroth/first\;power\;of\;R_2^{B,f}\;\;.
\eea
The corresponding operators of atom $B$ can be easily obtained by exchanging $A\rightleftharpoons B$ in the above expressions.

The expansion of the field operator, $\phi(x(\tau))$, is:
\beq
\phi(x(\tau))=\phi^{(0)}(x(\tau))+\phi^{(1)}(x(\tau))+\phi^{(2)}(x(\tau))+\phi^{(3)}(x(\tau))+O(\mu^4)
\eeq
with
\bea
\phi^{(0)}(x(\tau))=\phi^f(x(\tau))\;,\quad\quad\quad\quad\quad\quad\quad\quad\quad\quad\quad\quad\quad\;\;\\
\phi^{(1)}(x(\tau))=i\mu\sum_{\xi=A}^B\int_{\tau_0}^{\tau}d\tau_1R^{\xi,f}_{2}(\tau_1)[\phi^{f}(x_{\xi}(\tau_1)),\phi^{f}(x(\tau))]\;,
\label{first-order-field-operator}
\eea
\beq
\phi^{(2)}(x(\tau))=(i\mu)^2\sum_{\xi=A}^B\int_{\tau_0}^{\tau}d\tau_1\int_{\tau_0}^{\tau_1}d\tau_2\phi^{f}(x_{\xi}(\tau_2))
[\phi^{f}(x_{\xi}(\tau_1)),\phi^{f}(x(\tau))][R^{{\xi},f}_{2}(\tau_2),R^{{\xi},f}_2(\tau_1)]\;,
\eeq
and
\bea
\phi^{(3)}(x(\tau))&=&(i\mu)^3\int^{\tau}_{\tau_0}d\tau_1\int^{\tau_1}_{\tau_0}d\tau_2\int^{\tau_1}_{\tau_0}d\tau_3
[\phi^f(x_B(\tau_2)),\phi^f(x_A(\tau_1))][\phi^f(x_B(\tau_3)),\phi^f(x(\tau))]R_2^{A,f}(\tau_1)[R_2^{B,f}(\tau_2),R_2^{B,f}(\tau_3)]\nn\\&&
+(i\mu)^3\int^{\tau}_{\tau_0}d\tau_1\int^{\tau_1}_{\tau_0}d\tau_2\int^{\tau_2}_{\tau_0}d\tau_3
[\phi^f(x_B(\tau_2)),\phi^f(x_A(\tau_1))][\phi^f(x_B(\tau_3)),\phi^f(x(\tau))]R_2^{A,f}(\tau_1)[R_2^{B,f}(\tau_3),R_2^{B,f}(\tau_2)]\nn\\&&
+(i\mu)^3\int^{\tau}_{\tau_0}d\tau_1\int^{\tau_1}_{\tau_0}d\tau_2\int^{\tau_2}_{\tau_0}d\tau_3
[\phi^f(x_A(\tau_1)),\phi^f(x_B(\tau_3))][\phi^f(x_B(\tau_2)),\phi^f(x(\tau))]R_2^{A,f}(\tau_1)[R_2^{B,f}(\tau_3),R_2^{B,f}(\tau_2)]\nn\\&&
+(i\mu)^3\int^{\tau}_{\tau_0}d\tau_1\int^{\tau_1}_{\tau_0}d\tau_2\int^{\tau_2}_{\tau_0}d\tau_3
[\phi^f(x_A(\tau_3)),\phi^f(x_B(\tau_2))][\phi^f(x_B(\tau_1)),\phi^f(x(\tau))]R_2^{A,f}(\tau_3)[R_2^{B,f}(\tau_2),R_2^{B,f}(\tau_1)]\nn\\
&&+terms\;of\;the\;zeroth/first/third\;power\;of\;R_2^{B,f}\;\;.
\eea
Explicit expressions of the terms of the zeroth/first/third power in $R_2^{B,f}$ are not listed out as they are extremely tedious and do not contribute to the interatomic potential.

\end{widetext}

\end{document}